\documentclass[superscriptaddress,secnumarabic,
amssymb,amsmath,nobibnotes,aps,prd,showkeys,showpacs,nofootinbib]{revtex4}%
\usepackage{graphicx}
\usepackage{epsf}
\usepackage{epsfig}
\usepackage{scrtime}
\usepackage{multirow}
\usepackage[multiple]{footmisc}
\usepackage{amsfonts, graphics}
\usepackage{bm}
\usepackage[utf8x]{inputenc}
\usepackage{amsmath}
\usepackage{amsfonts}
\usepackage{amssymb}
\usepackage{color}
\usepackage{epstopdf}
\usepackage{natbib}%
\usepackage{multirow}
\usepackage{array}
\usepackage{hhline}
\usepackage{wrapfig}
\usepackage{lipsum}
\providecommand{\U}[1]{\protect\rule{.1in}{.1in}}

\begin{document}
\title{Mimicking the $\Lambda$CDM Universe through inhomogeneous space-time }
\author{Subhra Bhattacharya}
\email{subhra.maths@presiuniv.ac.in}
\affiliation{Department of Mathematics, Presidency University, Kolkata-700073, India}

\keywords{$\Lambda$CDM, Dark Matter, Dark Energy, Inhomogeneous space-time, Curvature}
\pacs{}

\begin{abstract}

Starting from an inhomogeneous space-time model of the universe we could recreate a scenario of recent time accelerating universe dominated by Dark Energy type of fluid. The background matter component of such a universe was considered to be made up of a combination of an anisotropic fluid, a barotropic fluid and the presureless cold dark matter. It was found that inhomogeneity exhibits itself as the curvature term in such a universe. We corroborated our model with recent supernova Ia-JLA data together with $H_{0}$ data and BAO data. Cosmographic analysis of the dynamical variables further show that the model can mimic the $\Lambda$CDM cosmology very closely.
\end{abstract}

\vspace{1em}

\maketitle

\vspace{1em}

\section{Introduction}

In the late nineties scientific perception of the evolving universe underwent a massive change as distance data from Type Ia Supernovae came in from the Supernova search team \cite{sup1} and Supernova cosmology project \cite{sup2}. It was found that the recent time observable universe was undergoing an accelerated rate of expansion driven by some unknown matter component with negative pressure. Dubbed as the Dark Energy (DE), the exact nature of this matter is one of biggest unresolved puzzles of modern day cosmology. Observation now predicts that about 70\% of the universe is made up of DE, while 27\% of is made up of pressure less cold dark matter (CDM). The remaining 3-4\% is the physically observable universe. Evidently determining the nature of the dark components of the universe, namely the DE and the Dark Matter (DM) is one of the thriving areas of current research. The most common and widely accepted explanation for DE is the cosmological constant $\Lambda.$ The cosmological constant together with CDM gives the $\Lambda$CDM model or the concordance model of the expanding universe. This is the most widely accepted and favoured description of the prevailing accelerating universe. However the model is not without its flaws. The two most severe problems being the coincidence problem and the cosmological constant problem. These together with the abstract definition for DE has forced researchers to look for other substantive explanations for DE. In fact there is no dearth of literature on viable descriptions for DE that range from dynamical DE models to massive gravity and modified gravity theories. (See \cite{amm} for a review of various DE models that exists in the literature). Despite the bulk of literature dedicated in identifying a physically viable DE model, none has yet been found.

Most of the models in normal gravity accept the Friedmann-Lemaitre-Robertson-Walker (FLRW) line element as the background geometry. This is due to the general assumption of a homogeneous and isotropic universe at large scales. However it is common knowledge that the observable universe at small scales is patchy and inhomogeneous. With the motivation of attending this question we start from a universe that has an inhomogeneous geometry. Accordingly we shall describe the evolution of the universe in the background of an inhomogeneous metric which will be asymptotically FRW. The spherically symmetric inhomogeneous line element used in our corresponding discussions has been widely used in the description of astrophysical objects called wormholes \cite{wrm}. In recent past such metric has also been used to describe emergent universe scenarios from different perspectives \cite{subh1,subh2}. We shall consider the corresponding physical component of the space-time to be made up of three non interacting fluid components. A pressure less CDM, a homogeneous and isotropic fluid with barotropic equation of state and another inhomogeneous and anisotropic fluid. Using the Friedmann's equations we shall derive the relevant mathematical expressions for the matter components. The model will be then subjected to numerical tests using the recent Supernova Ia Joint Light-Curve Analysis data (JLA), the Hubble data (OHD) and Baryon Acoustic Oscillations data (BAO). We shall also take a kinematic model independent approach to test the model parameters by running a cosmographic analysis and a $Om$ diagnostic test. Finally we shall test the performance of our model with respect to the $\Lambda$CDM model using the statistical model selection Akaike Information Criterion (AIC) and Bayesian Information Criterion (BIC).

The manuscript shall be organised as follows: In section 2 we shall describe the metric along with relevant gravitational and continuity equations with the corresponding analytical solutions. Section 3 shall contain an elaborate description of the numerical technique, data sets used and the numerical results. In section 4, 5 and 6 we shall present the model independent geometric cosmographic analysis, $Om$ test results and the results for Bayesian statistical model performance, that is the $AIC$ and $BIC$ criterion. Finally, section 7 gives a brief discussion on the results of the study.

\section{Model Description}

We consider a line element for inhomogeneous spherically symmetric space-time is given by:
 
\begin{equation}
ds^{2}=-dt^{2}+a^{2}(t)\left[\frac{dr^{2}}{1-b(r)}+r^{2}d\Omega^{2}\right].
\end{equation}
with $d\Omega^{2}=d\theta^{2}+\sin^{2}\theta d\phi^{2},~b(r)$ some arbitrary function of the radial component and $a(t)$ the scale factor of the universe.
The physical matter is considered to be composed three non interacting components as follows:
\begin{enumerate}
\item M1: A pressure less Cold Dark Matter with stress energy tensor denoted by $\rho_{m}(t)$
\item M2: A homogeneous and isotropic matter with constant equation of state $\omega$ given by: $p(t)=\omega\rho(t),~\rho~\text{and}~p$ being the energy and pressure components of the homogeneous matter tensor.
\item M3: An anisotropic inhomogeneous fluid having stress energy tensors $\xi(r,t)$ with radial and transverse pressure components $\eta(r,t)$ and $\tau(r,t).$
\end{enumerate}
The matter distribution corresponding to the M1 and M2 is given by:
\begin{equation}
T_{\mu\nu}=(p+\rho)u_{\mu}u_{\nu}+pg_{\mu\nu}\label{f12}
\end{equation}
(with corresponding $p=0$ adaptation for M1), while for M3 the corresponding matter distribution follows:
 \begin{equation}
 T_{\mu\nu}=(\xi+\tau)v_{\mu}v_{\nu}+\tau g_{\mu\nu}+(\eta-\tau)x_{\mu}x_{\nu}\label{f3}
 \end{equation}
Here $v_{\mu}$ and $x_{\mu}$ are unit time-like and space-like vectors respectively, satisfying
\begin{equation}
v_{\mu}v^{\mu}=-x_{\mu}x^{\mu}=-1,~x^{\mu}v_{\mu}=0.
\end{equation}
Accordingly the Einstein's field equations are given by:
\begin{align}
 H^{2}+\frac{b+rb^{'}}{3a^{2}r^{2}}&=\frac{8\pi G}{3}\left(\rho_{m}+\rho+\xi_{}\right)\label{field1}\\
 -(2\dot{H}+3H^{2})-\frac{b}{a^{2}r^{2}}&=8\pi G\left(p+\eta\right)\label{field2}\\
 -(2\dot{H}+3H^{2})-\frac{b^{'}}{2a^{2}r}&=8\pi G(p+\tau)\label{field3}
 \end{align}
 where $H=\frac{\dot{a}}{a}$ is the Hubble parameter with over-dot denoting differentiation with respect to cosmic time $t$ while prime indicates differentiation with respect to the radial component $r.$ Since we have considered a scenario of non-interacting fluids the corresponding energy conservation equations are given by:
  \begin{align}
  \dot{\rho_{m}}+3H\rho_{m}&=0\label{econsv0}\\
 \dot{\rho}+3H\rho(1+\omega)&=0 \label{econsv1}\\
 \frac{\partial\xi}{\partial t}+H(3\xi+\eta+\tau)&=0\label{econsv2}\\
 \frac{\partial \eta}{\partial r}=\frac{2}{r}(\tau-\eta)\label{econsv3}
 \end{align}
Considering that the stress energy tensor of the anisotropic fluid is related to the corresponding radial and transverse pressure by a barotropic equation of state given by: $\eta=\omega_{r}\xi$ and $\tau=\omega_{t}\xi$ we can explicitly obtain the corresponding anisotropic stress tensor as \cite{subh1}:
 \begin{equation}
 \xi(t,r)=\xi_{0}\frac{r^\frac{2(\omega_{t}-\omega_{r})}{\omega_{r}}}{a^{3+\omega_{r}+2\omega_{t}}}\label{ed1}
 \end{equation}
with $\xi_{0}$ is the constant of integration. Accordingly for $\omega_{r}+2\omega_{t}+1=0$ the functions $b(r), \xi(r,t),~\eta(r,t)~\text{and}~\tau(r,t)$ are explicitly solved as
\begin{equation}
 b(r)=b_{0}r^{2}-8\pi G\xi_{0}\omega_{r}r^{-(\frac{1}{\omega_{r}}+1)}\label{b}
 \end{equation}
 \begin{equation}
\xi(t,r)=\frac{\xi_{0}}{a^{2}}r^{-(3+\frac{1}{\omega_{r}})}\qquad  \eta(r,t)=\omega_{r}\xi\qquad ~\tau(r,t)=-\frac{1+\omega_{r}}{2}\xi.\label{p}
 \end{equation}
Further equations (\ref{econsv0}) and (\ref{econsv1}) give $\rho_{m}(t)=\rho_{m0}a^{-3}$ and $\rho(t)=\rho_{0}a^{-3(1+\omega)}.$ If the above obtained solutions are substituted in equation (\ref{field1}) we obtain:
\begin{equation}
H^{2}=-\frac{b_{0}}{3}a^{-2}+\frac{8\pi G}{3}\left(\rho_{m0}a^{-3}+\rho_{0}a^{-3(1+\omega)}\right)\label{field4}
\end{equation}
Evidently the anisotropic matter component expresses itself as the spatial curvature of the universe which was inherent in the space-time due to the inhomogeneity and anisotropy assumption. Our model thus, accounts for the spatial curvature of the universe without any prior assumption on the same. From equation (\ref{p}) one can observe that the anisotropic matter component is significant at smaller spatial scales, giving rise to local inhomogeneity and anisotropy. Whereas at large spatial scales the, corresponding to $\omega_{r}>-\frac{1}{3}$ the inhomogeneity vanishes resulting in the observable homogeneous and isotropic universe. It is significant that irrespective of the spatial scale the only contribution of this matter component to the energy budget of the universe is corresponding to an inherent spatial curvature. Thus we see that with the initial assumption of the inhomogeneous and anisotropic universe, the Friedmann equations essentially point to the existence of a curvature component.

\section{Data and Numerical Results}

The exact nature and significance of each fluid component present in the energy budget can be determined if we subject the model to numerical tests and constrain the free parameters $b_{0},~\omega,$ and $\rho_{0}.$ Data sets from type Ia supernovae JLA data, OHD data and BAO data is used for the analysis. Further, the significance of the DE like unknown matter component M2 can also be explained using the numerical tests. It will also help us determine the significance of the curvature term. Analytically we saw that the inhomogeneous term expressed itself as the curvature component similar to that in an FLRW line element of a homogeneous and isotropic universe. Using numerical results we will be able to establish the significance of the term.   

\subsection{Data Sets Used}

\begin{description}
\item {\it JLA Data}: We shall use a sample of 31 binned data sets from the Joint Light-curve analysis \cite{jla}. The luminosity distance $d_{L}(z),$ of a type Ia supernova located at red shift $z$ is related to its corresponding Hubble parameter in a flat universe by  the relation $d_{L}(z)=(1+z)\int_{0}^{z}\frac{1}{H(z')}dz'.$ In a universe with spatial curvature a corresponding modification has to be introduced to account for the curvature, accordingly we redefine $d_{L}(z)$ as follows:
\begin{equation}
d_{L}(z)=\frac{(1+z)}{H_{0}}Re\left(\frac{\sinh(H_{0}\sqrt{\Omega_{k}}\int_{0}^{z}\frac{1}{H(z')}dz')}{\sqrt{\Omega_{k}}}\right).\label{dl}
\end{equation}
This is related to the observed distance modulus $\mu(z)$ of the type Ia supernova at red shift $z$ by the relation:
\begin{equation}
\mu(z)=m_{B}-M_{B}=5\log_{10}\left(\frac{d_{L}(z)}{1Mpc}\right)+25
\end{equation}
where $m_{B}$ and $M_{B}$ are the corresponding apparent and absolute magnitude of a observed supernova Ia in the $B$ band. The corresponding $\chi^{2}$ function is given by:
\begin{equation}
\chi^{2}_{JLA}=\frac{1}{2}(\boldsymbol{\mu^{obs}}-\boldsymbol{\mu})^{T}\boldsymbol{C}_{\mu}^{-1}(\boldsymbol{\mu^{*}}-\boldsymbol{\mu})
\end{equation}
with $\boldsymbol{C}_{\mu}$ the covariance matrix for the binned JLA data set.

\item {\it OHD Data:} We shall use 35 sets of observational Hubble data corresponding to red shift range $0.07\leq z\leq 2.36.$ The data sets have been derived from Cosmic Chronometers \cite{cc}, the baryon acoustic oscillation method \cite{al} and the baryon acoustic oscillations in the Ly$\alpha$ forest of high red shift quasars \cite{del}. The corresponding $\chi^{2}$ is defined as:
\begin{equation}
\chi^{2}_{OHD}=\sum_{i=1}^{35}\frac{\left(H^{obs}(z_{i})-H(z_{i}(\Theta))\right)^{2}}{\sigma_{i}^{2}}
\end{equation}

\item {\it BAO Data:} We have used four sets of BAO data obtained from the following sources: 
\begin{enumerate}
\item 6dF Galaxy Survey at $z=0.106$ \cite{df},
\item the SDSS DR7 main galaxy sample at $z=0.15$  \cite{sd},
\item the BOSS LOWZ data at $z=0.32$ and
\item BOSS CMASS at $z=0.57$  \cite{boss}.
\end{enumerate}
The $\chi^{2}$ of the BAO data is given by:
\begin{equation}
\chi^{2}_{BAO}=\sum_{i=1}^{4}\frac{(\boldsymbol{(\eta}^{obs}(z_{i})-\boldsymbol{\eta}(z_{i}\Theta))^{2}}{\boldsymbol{\sigma}_{i}^{2}}
\end{equation}
where $\boldsymbol{\eta}(z_{i})$ is related to acoustic distance ratio $d_{V}/r_{s}$ with $r_{s},$ the co-moving size of sound horizon at baryon drag epoch and $d_{V},$ the dilation scale, which is given by $d_{V}=\left(\frac{zd_{L}(z)}{H(z)}\right)^{\frac{1}{3}}.$

\end{description}

\subsection{Statistical Tool Used}
We use the Markov Chain Monte Carlo (MCMC) technique of parameter estimation using the $Python$ application of the MCMC devised by Goodman and Weare \cite{gd} and Foreman-Mackey {\it et al} \cite{fr}. The background statistical analysis is based on the Bayesian approach based on determining the posterior probability distribution function of the parameter space. By the Bayes' theorem we can relate the posterior probability distribution of the parameter space to the prior probability distribution of the parameters and the corresponding likelihood function \cite{nil}. Accordingly the minimum $\chi^{2}$ or the maximum likelihood function is given as follows:
\begin{equation}
-\ln \mathfrak{L}(\Theta)=\frac{1}{2}\left(\chi^{2}_{JLA}+\chi^{2}_{OHD}+\chi^{2}_{BAO}\right)
\end{equation}
where $\Theta$ is the set of parameters.

\subsection{Numerical Results}

\begin{table}
\begin{tabular}{c|c|c|c|c|c}

\hhline{======}
Data&$\chi^{2}_{min}/d.o.f$&$h_{0}$&$\Omega_{k}$&$\Omega$&$\omega$\\
\hline
JLA+OHD&$~~0.807~~$&$0.7^{+0.01}_{-0.01}$&$0.06^{+0.19}_{-0.20}$&$0.7^{+0.16}_{-0.15}$&$-0.95^{+0.12}_{-0.18}$\\
\hline
JLA+OHD+BAO&$~~0.802~~$&$0.7^{+0.01}_{-0.01}$&$0.06^{+0.21}_{-0.20}$&$0.7^{+0.16}_{-0.15}$&$-0.93^{+0.11}_{-0.18}$\\
\hhline{======}
\end{tabular}
\caption[One]{First table}
\end{table}

We have constrained our parameters using JLA+OHD data and JLA+OHD+BAO data. In Figure 1 and 2 we have plotted the $1\sigma$ (68.3\%), $2\sigma$ (95.4\%) and $3\sigma$ (99.7\%) confidence contours corresponding to the constrained parameters $h_{0}$ the rescaled Hubble parameter with $h_{0}=H/100 km~sec^{-1}Mpc^{-1},$ the equation of state $\omega$ for M2, the rescaled density parameter $\Omega$ given by $\Omega=\frac{\rho_{0}}{\rho_{c}}$ where $\rho_{c}=\frac{3H^{2}}{8\pi G}$ and $\Omega_{k}=-\frac{b_{0}}{3H^{2}}.$ Figure 1 shows the confidence contours with JLA+OHD data and Figure 2 shows the confidence contours for JLA+OHD+BAO data sets. Both figures also show the marginalized posterior distribution of the parameters with $\pm~1~\sigma$ error bars. 

Table 1 is a tabular display of the results obtained corresponding to the two data sets used in the analysis together with their $\pm~1~\sigma$ error bar. We have also provided the $\chi^{2}$ value corresponding to the number of degrees of freedom. 

From the results it is evident that the homogeneous matter density for M2 behaves as a dark energy fluid with the constrained values for the equation of state parameter $\omega\equiv -0.94$ and the rescaled density parameter $\Omega=0.70$ Further the parameter $\Omega_{k},$ which was interpreted analytically as the curvature density is constrained to $\Omega_{k}=0.06$ The rescaled Hubble parameter $h_{0}$ was also constrained to $h_{0}=0.70$ The numerical results show that the parameters closely follow the $\Lambda$CDM limit. Here the matter M2 is synonymous with the dark energy density, while the curvature density is indicative of the existence of a small curvature. From the equation of state parameter for $\omega$ we see that although it has a value close to cosmological constant, yet it has not crossed the phantom divide barrier. 
\begin{figure}[htb]
\begin{center}
\includegraphics[width=13cm,height=9cm]{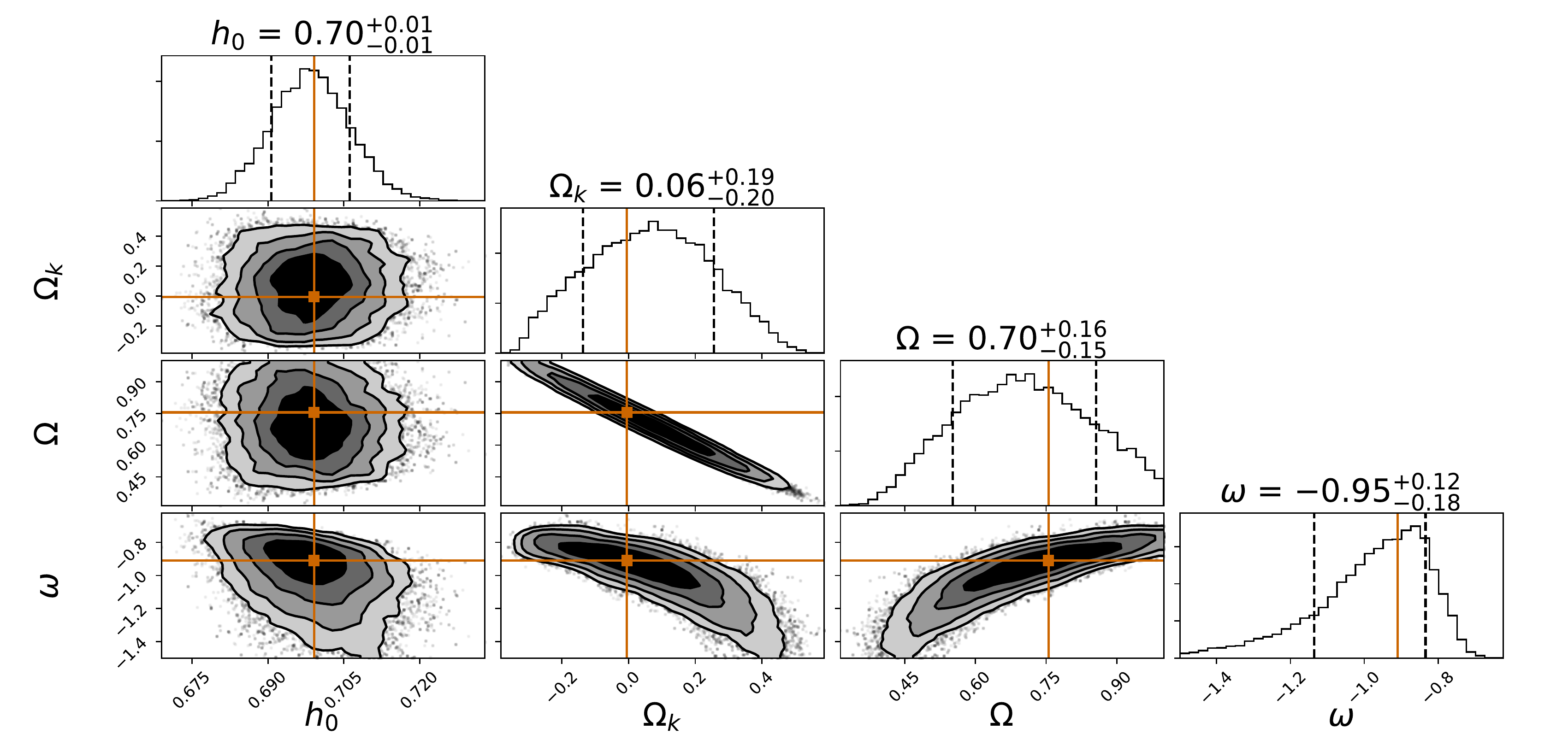}
\caption{The above figures show the confidence contour on a two dimensional parameter space with the marginalised posterior distributions of $(h_{0},\Omega_{k},\Omega,\omega)$ obtained by a combined analysis of JLA and OHD data. The median values of the constrained parameters with corresponding $1\sigma$ errors are given on top of the posterior probability distribution panels.}
\end{center}
\end{figure}

\begin{figure}[htb]
\begin{center}
\includegraphics[width=13cm,height=9cm]{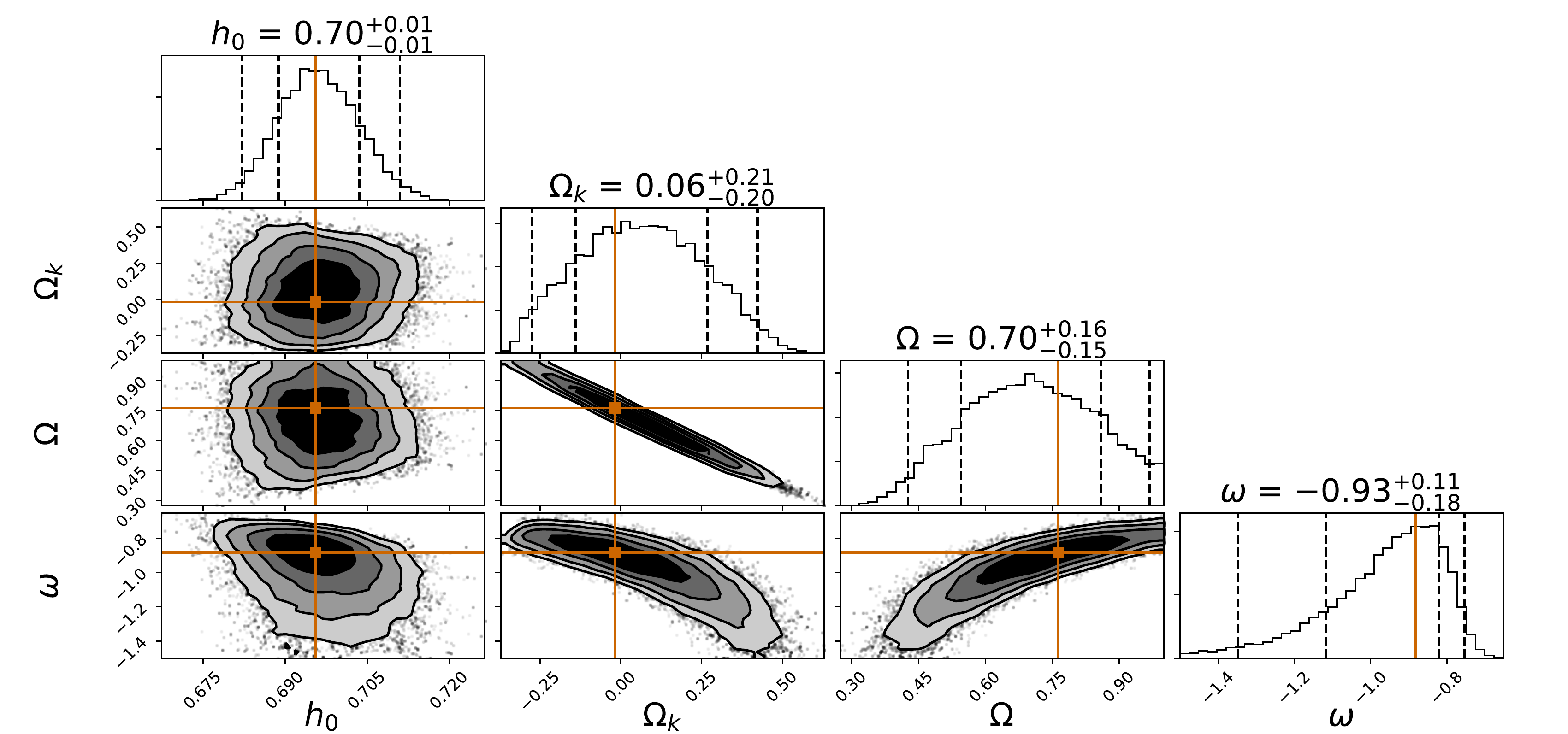}
\caption{This figure show the confidence contour of the marginalised posterior distributions of $(h_{0},\Omega_{k},\Omega,\omega)$ obtained from a combined analysis of JLA+OHD+BAO data. The median values of the constrained parameters with corresponding $1\sigma$ errors are given on top of the posterior probability distribution panels. }
\end{center}
\end{figure}

\section{Cosmographic Analysis} 

As a next step, we use two model independent geometric diagnostic to characterize the model. One is the cosmographic diagnostic which was introduced in \cite{sh} by the name of Statefinder parameters, to tackle the degeneracy among various DE models. Initially two Statefinder parameters $r,s$ were introduced. Later Visser \cite{vb} introduced other cosmographic parameters by considering the higher order terms appearing in the Taylor series expansion of the scale factor $a(t)$ about the present time $t_{0}$. These parameters, obtained as the derivative of the scale factor $a(t),$ namely the Hubble, deceleration, jerk and snap parameters is defined as follows:
\begin{align*}
H(t)=\frac{1}{a}\frac{da}{dt}\qquad q(t)=-\frac{1}{aH^{2}}\frac{d^{2}a}{dt^{2}}\\
j(t)=\frac{1}{aH^{3}}\frac{d^{3}a}{dt^{3}}\qquad s(t)=-\frac{1}{aH^{4}}\frac{d^{4}a}{dt^{4}}
\end{align*}
(the parameter $r$ introduced in \cite{sh} is same as the parameter $j$ by \cite{vb}, however the snap in \cite{vb} is different from the $s$ in \cite{sh}.)
 
Accordingly all these parameters can be represented in terms of the deceleration parameter $q$ as follows:
\begin{equation}
j=(1+z)\frac{dq}{dz}+q(1+2q)\qquad s=-(1+z)\frac{dj}{dz}+j-3j(2+q).
\end{equation} 
with the deceleration parameter $q$ expressed in terms of the Hubble parameter by the relation $q=-1-\frac{\dot{H}}{H^{2}}.$ Using our model parameters we show the evolution of the cosmographic $q,~j$ and $s$ parameter. Figure 3 shows how the cosmographic parameters evolve for the model w.r.t. the standard $\Lambda$CDM model. From the figure we see that the model performs quite well w.r.t the standard concordance model. The deceleration parameter gives an early decelerating universe followed by a late time accelerating universe. The transition from deceleration to acceleration happening around $z=0.65.$  The results being consistent in both JLA+OHD and JLA+OHD+BAO. The $j$ parameter maintains an approximately constant value of 1 only in early times. In late time it ranges between 0.8 to 0.9. The $s$ parameter too attains values close to that of $\Lambda$CDM model. The cosmographic parameters further validates that the model has strong affinity towards the concordance model of expanding universe.

\begin{figure}[htb]
\begin{center}
\includegraphics[width=5cm,height=4cm]{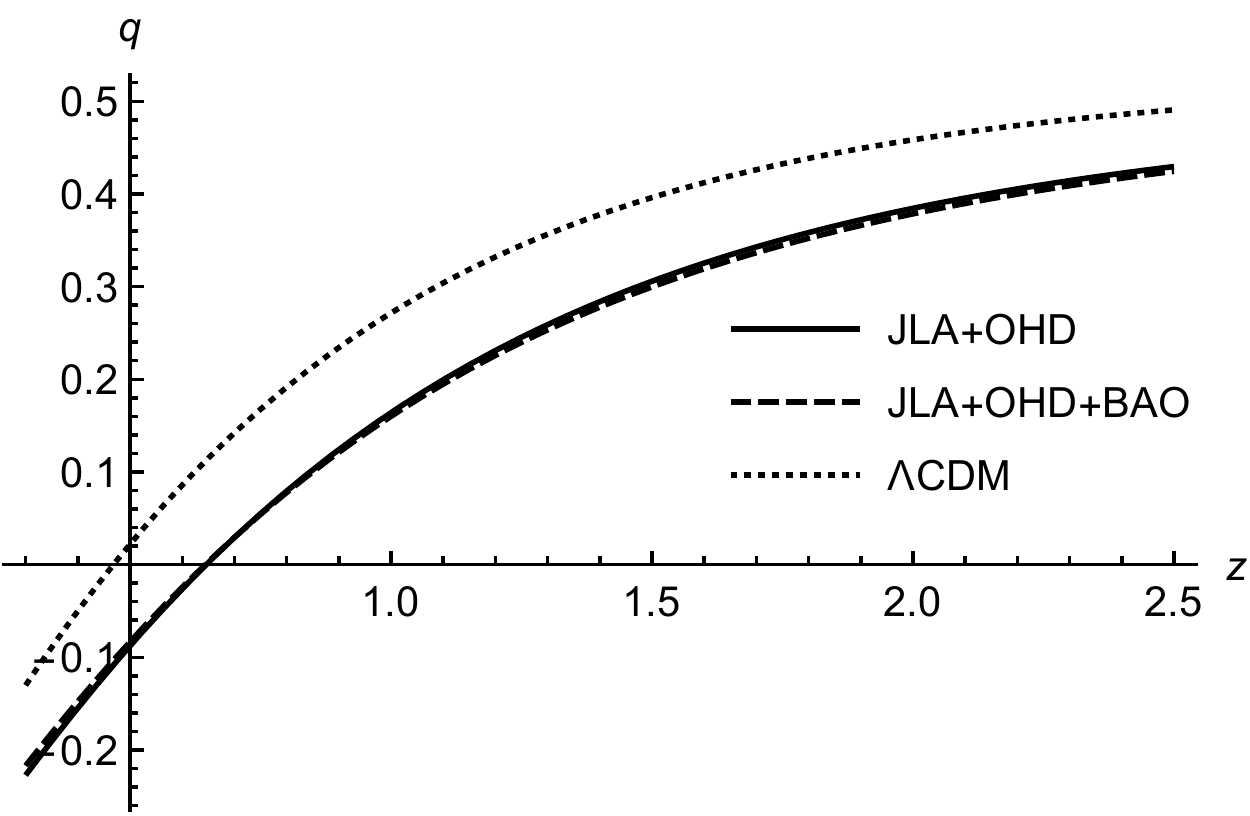}\quad\quad \includegraphics[width=5cm,height=4cm]{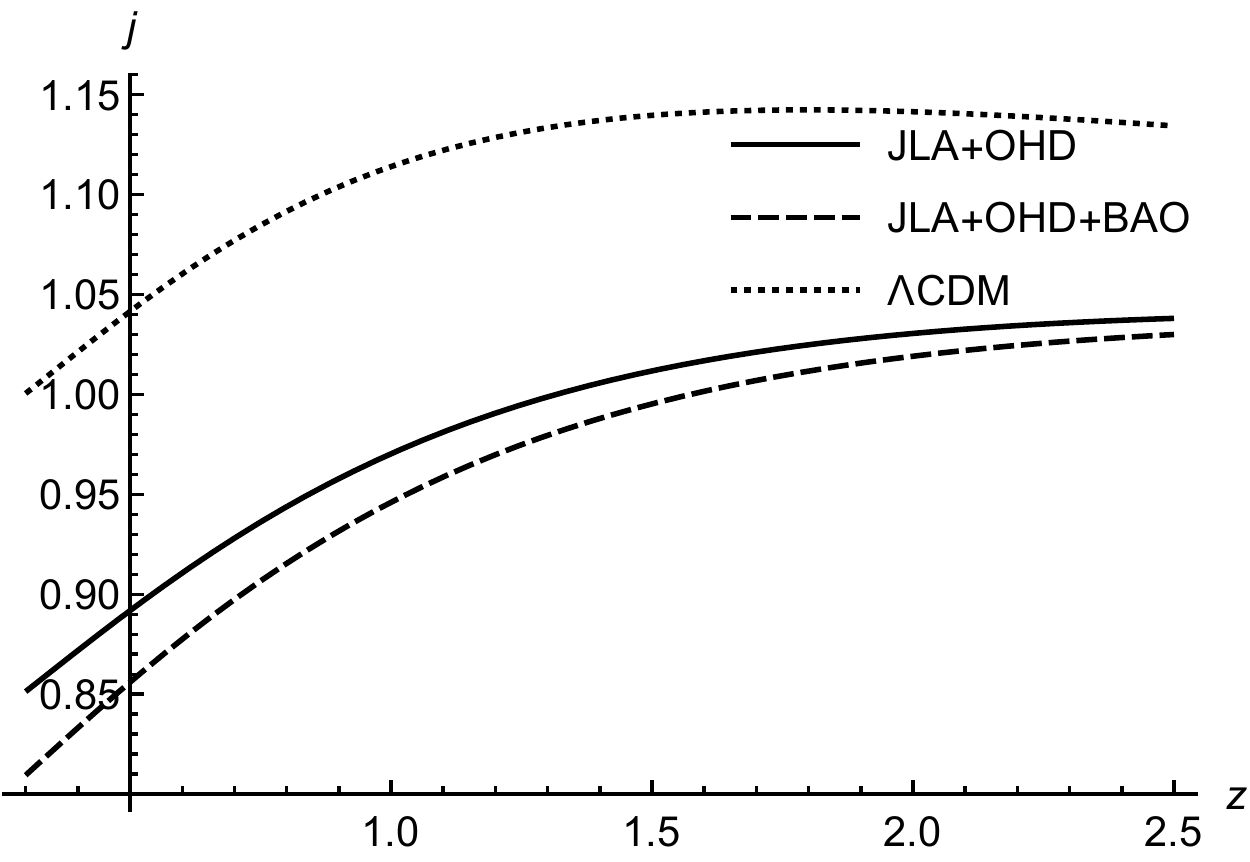}\quad\quad
\includegraphics[width=5cm,height=4cm]{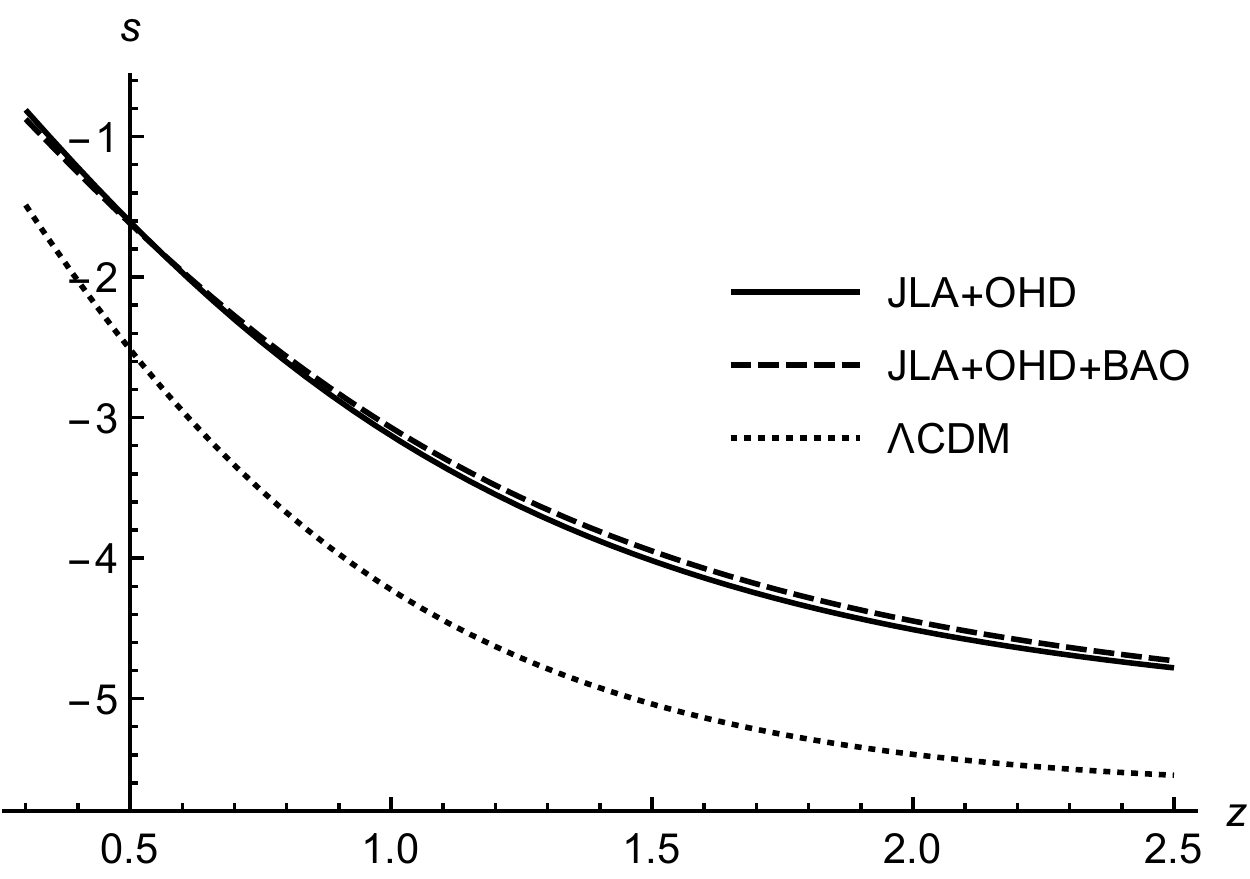}
\caption{The figures show how the cosmographic deceleration parameter $q,$ the jerk parameter $j$ and the snap parameter $s$ obtained corresponding to the model parameters as constrained by JLA+OHD data and JLA+OHD+BAO data compare w.r.t the standard $\Lambda$CDM model.}
\end{center}
\end{figure}

\section{The $Om$ Test}

The $Om$ diagnostic test, is another model independent test which was introduced in \cite{om}. For $\Lambda$CDM model with cosmological constant, $Om$ is defined as $Om(z)=\frac{h^{2}(z)-1}{(1+z)^3-1}.$ This is again a test of deviation from $\Lambda$CDM cosmology. Ideally for $\Lambda$CDM  with dark energy density -1, we get $Om(z)=\Omega_{m},$ that is it remains constant at the matter density scaled by the critical density. For any other dark energy with equation of state $\omega$ we have $Om(z)=\Omega_{m}+(1-\Omega_{m})\left[\frac{(1+z)^{3(1+\omega)}-1}{(1+z)^{3}-1}\right].$ Accordingly the difference between $Om(z)$ for $\Lambda$CDM, with cosmological constant tested for two different red shifts should ideally be zero. For models with quintessence dark energy ($\omega>-1$) we should ideally get $Om(z)>\Omega_{m}$ and for phantom dark energy models ($\omega<-1$) $Om(z)<\Omega_{m}.$ Thus the difference between $Om(z)$ for the model and $\Lambda$CDM gives us an idea of the model in a model independent way. 

In figure 4 we plot the difference between $Om(z)$ and $\Omega_{m}$ corresponding to our results as obtained with JLA+OHD and JLA+OHD+BAO data. The plot show the difference between the $Om(z)$ of the model and the $\Omega_{m}$ of $\Lambda$CDM model. Since $Om(z)>\Omega_{m}$ the results point to a scenario of the quintessence dark energy.

\begin{wrapfigure}[]{R}{6.5cm}
\begin{center}
\includegraphics[width=6cm,height=5cm]{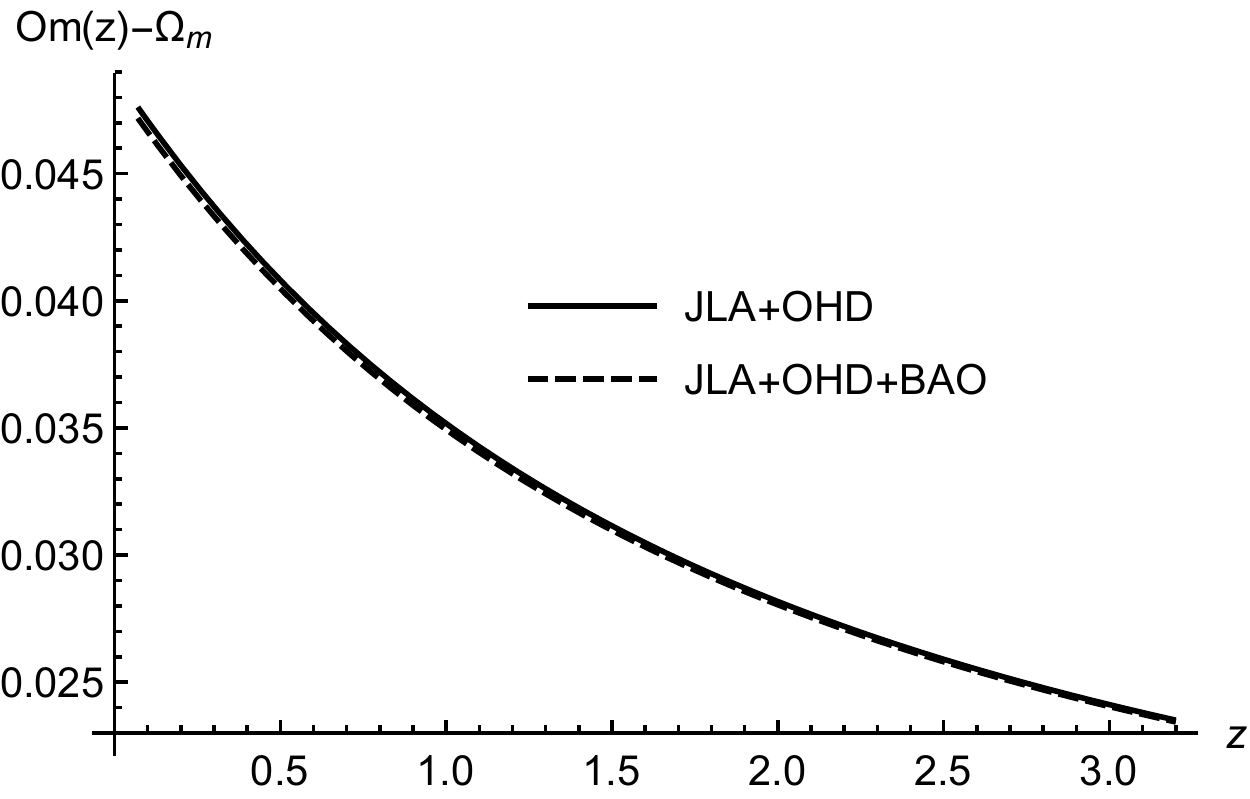}
\caption{The figure shows the values of $Om(z)$ corresponding to the parameters as obtained with two data sets used}
\end{center}
\vspace{-20pt}
\end{wrapfigure}
  
\section{Statistical Model Selection}

Distinguishing a model based on its performance with respect to a set of other competing models can be done using several statistical analysis. Among the several such statistical techniques available two have is been used more often than others \cite{lid}. One is the Akaike Information Criterion (AIC) based on information theory, while the other is the Bayesian Information Criterion (BIC) is based on Bayesian evidence. The AIC was introduced by Akaike in 1974 \cite{ak}, while BIC by Schwarz in 1978 \cite{bk}. Both can be applied to cosmological models based on knowledge of the maximum likelihood estimates of the model.
The AIC is defined as 
\begin{equation}
AIC=-2\ln\mathfrak{L}_{max}+2n
\end{equation} where $n$ is the number of parameters of the model. The best model minimizes AIC. The BIC is defined as 
\begin{equation}
BIC=-2\ln\mathfrak{L}_{max}+n\ln N 
\end{equation}
where $N$ is the number of data points used to constrain the model. In case of small data sets another modified version of AIC given by \cite{sg}
\begin{equation}
AIC_{c}=AIC+\frac{2n(n+1)}{N-n-1}.
\end{equation} 

Evidently for $N\gg n$ the correction introduced looses significance and hence $AIC_{c}\approx AIC.$ For all other cases it was stressed in \cite{burn} that $AIC_{c}$ works better compared to the previous version. Significance of the model is then judged based on the difference of the AIC and BIC values with respect to some competing model. The scale of evaluation is based on ``Jeffry's" index scale where a $\Delta IC>5$ is considered to be strong evidence against the model, while a $\Delta IC>10$ provides decisive evidence against the model. Table II provides the values of $\Delta IC$ corresponding to the model with parameters as constrained by the data sets. From the table we see that the $\Delta IC$ values are all $<5.$ This shows that the model is significant and comparable to the standard $\Lambda$CDM.

\begin{table}
\begin{tabular}{c|c|c|c}
\hhline{====}
Data&$\Delta AIC$&$\Delta AIC_{c}$&$\Delta BIC$ \vspace{-1em}\\
\hline
JLA+OHD&$~~1.75~~$&$~~2.02~~$&$~~3.94~~$\vspace{-1em}\\
\hline
JLA+OHD+BAO&$~~1.53~~$&$~~1.78~~$&$~~3.78~~$\vspace{-1em}\\
\hhline{====}
\end{tabular}
\caption[One]{The difference in $AIC$, $BIC$ and $AIC_{c}$ of the model when compared to a $\Lambda$CDM model. }
\end{table}
\section{Discussion}

It is now an established fact that the universe is accelerating its expansion. The mystery to this is however the agent that drives the acceleration. The most well known theory that explains this expansion is the $\Lambda$CDM model, a model of a flat universe dominated by the cosmological constant $\Lambda$ and cold dark matter. In this work we have tried to put a different perspective of the same model by taking into consideration the local inhomogeneities present in the universe. Thus starting with a inhomogeneous universe we could analytically account for a tiny non-zero positive curvature of the universe. Using analytical techniques we obtained a universe that is homogeneous and isotropic at large scales with inhomogeneities remaining significant at small spatial scales. From the Friedman equations and the continuity equations of the fluid we could find explicit expression of the Hubble parameter $H.$ The Hubble parameter had an isotropic matter component, a curvature like component and the cold dark matter component. Using data from 1a Supernovae, Hubble data and BAO data we could constrain our unknown parameters that revealed a universe with a tiny positive curvature dominated by a quintessence type DE component. Further model independent tests revealed the strong tendency of the model to mimic the $\Lambda$CDM cosmology. 

Essentially we have provided an alternate view of the $\Lambda$CDM universe. Interestingly the initial assumption of the inhomogeneous universe finally accounted for a tiny curvature with dark energy DE matter component being quintessence type rather than the cosmological constant.

\section{ Acknowledgemensts:}

SB acknowledges UGC's Faculty Recharge Programme and Department of Science and Technology, SERB, India for financial help through ECR project (File No. ECR/2017/000569). The author acknowledges Mr. Niladri Paul for useful discussions and assistance with the $Python$ programming. 
SB also acknowledges IUCAA, Pune, India, for their hospitality while working on this project.

\end{document}